\documentclass[a4paper]{article}

\usepackage{icrc2013}

\title{The ASTRI Mini-Array Software System}

\shorttitle{The ASTRI Mini-Array Software System}

\authors{ Gino Tosti$^{1}$, Joseph Schwarz$^{2}$, Lucio Angelo Antonelli$^{3}$, Massimo Trifoglio$^{4}$,  Giuseppe Leto$^{5}$, Fulvio Gianotti$^{4}$, Rodolfo Canestrari$^{2}$, Osvaldo Catalano$^{6}$, Mauro Fiorini$^{7}$, Enrico Giro$^{8}$, Nicola La Palombara$^{3}$, Maria Concetta Maccarone$^{6}$, Giovanni Pareschi$^{2}$, Luca Stringhetti$^{7}$, and Stefano Vercellone$^{6}$ on behalf of the ASTRI collaboration$^{9}$ }

\afiliations{
$^1$Universit\`a di Perugia\\
$^2$INAF-Osservatorio Astronomico di Brera\\
$^3$INAF-Osservatorio Astronomico di Roma\\
$^4$INAF-Istituto di Astrofisica Spaziale e Fisica Cosmica di Bologna\\
$^5$INAF-Osservatorio Astrofisico di Catania\\
$^6$INAF-Istituto di Astrofisica Spaziale e Fisica Cosmica di Palermo\\
$^7$INAF-Istituto di Astrofisica Spaziale e Fisica Cosmica di Milano\\
$^8$INAF-Osservatorio Astronomico di Padova\\
$^9$http://www.brera.inaf.it/astri/}

\email{gino.tosti@fisica.unipg.it}

\abstract{ASTRI (``Astrofisica con Specchi a Tecnologia Replicante Italiana") is a Flagship Project financed by the Italian Ministry of Education, University and Research, and led by INAF, the Italian National Institute of Astrophysics. Main goals of the ASTRI project are the realization of an end-to-end prototype of a Small Size Telescope (SST) for the Cherenkov Telescope Array (CTA) in a dual-mirror configuration (SST-2M)
and, subsequently, of a mini-array composed of a few SST-2M telescopes to be placed at the final CTA Southern Site. Here we present the main features of the Mini-Array Software System (MASS) that  has a central role in the success of the ASTRI Project and will also serve as a prototype for the CTA software system. The MASS will provide a set of tools to prepare an observing proposal, to perform the observations specified therein (monitoring and controlling all the hardware components of each telescope), to analyze the acquired data online and to store/retrieve all the data products to/from the archive.}

\keywords{High-energy astrophysics, Cherenkov Telescope, CTA, ASTRI, Telescope System Software }

\begin{document}
\maketitle

\section{Introduction}
ASTRI is a {\it flagship} program financed by the Italian Ministry of Education, University and Research (MIUR) to develop special technologies suitable for the ambitious CTA Observatory~\cite{cta}. The CTA is composed by many tens of telescopes divided in three kinds of classes, in order to cover the energy range from a tens of GeV (Large Size Telescope, LST), to a tens of TeV (Medium Size Telescope, MST), and up to 100 TeV and beyond (Small Size Telescope, SST)~\cite{ssticrc}. Within this framework, INAF, the Italian National Institute of Astrophysics,  is currently~\cite{astrisys}:  developing an end-to-end prototype of SST in a dual-mirror configuration (SST-2M) to be tested under field conditions, and scheduled to start data acquisition in late 2014; the design and deployment of a mini-array of SST-2M as a first seed for the CTA Observatory. \\ 
In the framework of the ASTRI Project, the Mini-Array Software System (MASS) has the  task of making it possible to operate both  the ASTRI SST-2M and the Mini-Array under the  constraints and requirements contained in the ASTRI Operation Modes and User 
Requirements document. \\
The MASS is being designed to allow a member of the ASTRI Collaboration (the user) 
to perform all the steps needed to prepare and execute a scientific observing run. 
The MASS will provide a set of tools to the user  to prepare an observing proposal, 
to perform the observations specified therein, to analyze the acquired data online and to retrieve 
all the data products from the archive. \\
Also, the MASS will provide the software 
for the management and administration of proposals, scheduling of the observations, instrument  operations and use of the archive. 
Furthermore the design of the MASS must be compliant with the CTA 
Data Management (CTA-DM) and CTA Data Acquisition and Array Control  (CTA-ACTL) 
requirements and guidelines. \\
The  base unit of the ASTRI Mini-Array is the ASTRI SST-2M and in this paper we will describe the main characteristics of the software dedicated to control it. 
 \begin{figure}[lt]
  \centering
  \includegraphics[scale=0.6]{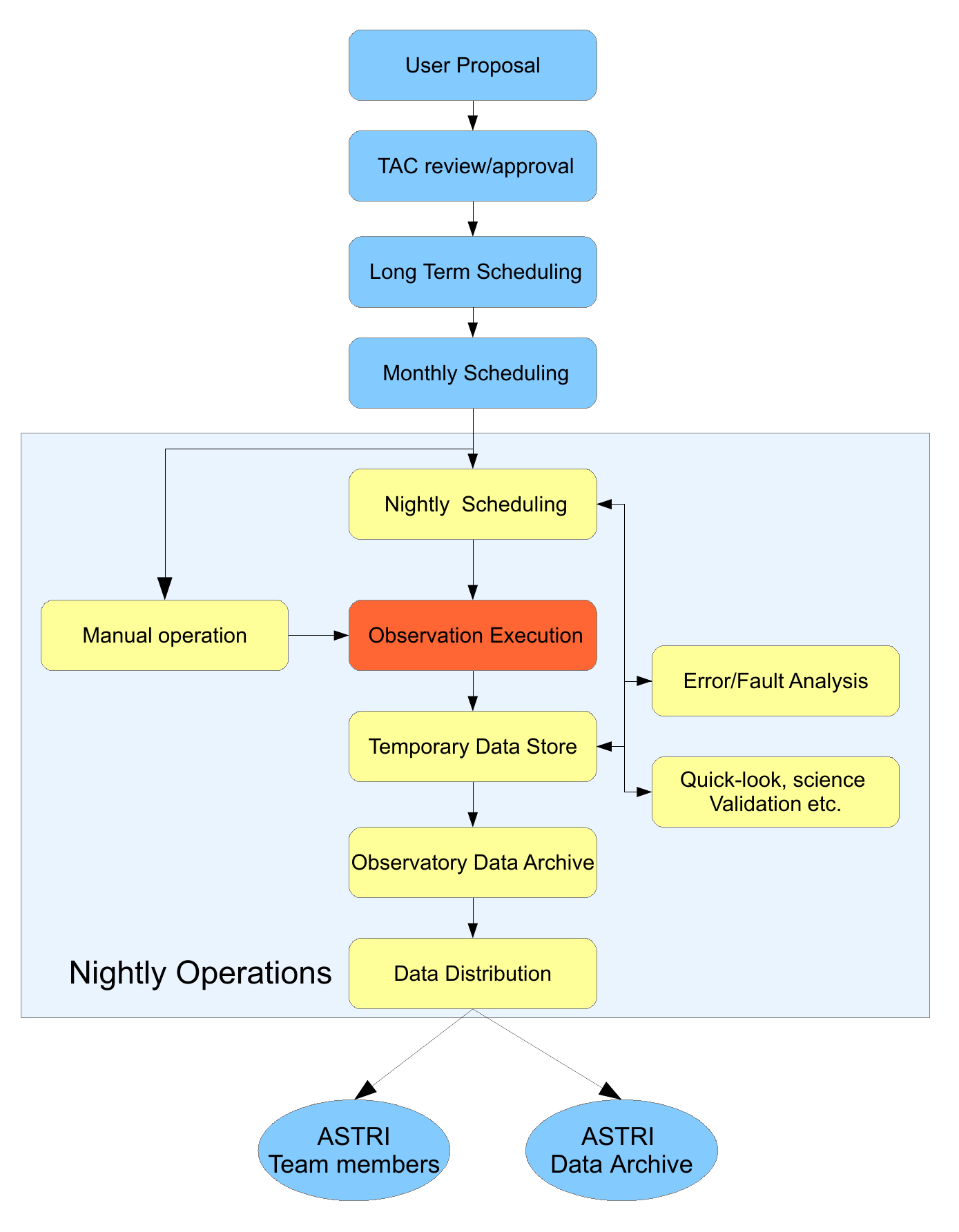}
  \caption{MASS information flow}\label{fig:fig1}
 \end{figure}

\section{THE ASTRI SST-2M Hardware Architecture of the Control  System\label{hw}}

The ASTRI SST-2M is a complex system including several subsystems, such as telescope,  camera and safety systems, that have to be monitored in order to get, in real time, the current status and the availability of each component. Also, the different  parts of the system have to be controlled in order to perform the operations requested  by the users (astronomers, operators, engineers, etc.).  A detailed description of the mechanical, kinematical and optical characteristics of the SST-2M can be found here \cite{astristruct}.\\
The ASTRI SST-2Mtelescope Azimuth and Elevation servo systems are based on brushless motors. Absolute encoders monitor the position of the axes. State of the art  PC-based technology is used to control the pointing and tracking of the  telescope; this eliminates the need for complex and expensive  special hardware (such as a Programmable Logic Controller), since the functionality (control, regulation, motion control and visualization) of this hardware can be handled by the software.  A fieldbus is used to connect the secondary servo systems, the Active Primary Mirror (M1) and the Secondary Mirror (M2) piston/tilt control system as well as other I/O devices. A large number 
of temperature sensors, strain-gauges and accelerometers are distributed along 
the telescope structures, mirrors and mirror supports.  Given the pointing requirements, 
$\pm10$ arcsec rms, up to $0.5^\circ$ of zenith angle, two monitoring systems are installed:
\begin{itemize}
\item One wide field CCD camera (field of view of  $\sim10^\circ $) mounted on the M1 support  to obtain an astrometric solution 
for the parallactic angle and also useful to derive and monitor the pointing model of  the telescope (sampling rate of the sky images about 1 Hz); 
\item One Giga Ethernet CCD camera mounted on the ASTRI camera and 18 laser diodes 
(one for each M1 dish panel) using direct telescope optics to monitor panel and 
secondary flexures due to wind and gravity (sampling rate for images about 10 Hz).
\end{itemize}
The ASTRI Camera Back-End is in charge of delivering the Camera bulk data to the 
Camera Server through a TCP/IP connection on a 1 (possibly 10) Gbit fiber Ethernet 
LAN.
The system includes several other auxiliary devices such as the time distribution 
system, which will be based on a GPS receiver and NTP server (with possible use 
of the CERN White Rabbit technology for the mini-array, but not the first telescope 
prototype). Furthermore, there are other devices dedicated to the site and environmental monitoring.

\section{ASTRI SST-2M and Mini-Array System Software design Concept}

The collection of sub-systems described in section ~\ref{hw}, must act in concert under 
the control of the MASS to achieve the goals stated in the ASTRI System Requirement 
Document (SRD) both for the single telescope and for the Mini-Array.  In particular 
the general operations that should be managed by the MASS software are given in 
the ASTRI Operation Procedures document and User Requirements document and are compliant with the CTA requirements. \\

The information flow that will be managed by the MASS is shown in Figure ~\ref{fig:fig1}. 
\begin{itemize}
\item The ASTRI Team user initiates the Observing cycle by preparing and submitting a 
proposal to the ASTRI Time Allocation Commitee (TAC). After the proposal has been reviewed and accepted it 
is inserted in the ASTRI Data Archive. The user observation is then included in 
the Observation plan and in the long term Operations Plan.  
\item The scheduling system will include the observation in the appropriate nightly schedule and it 
 will be performed by the MASS Observation Execution subsystem.  The 
raw data collected during the observation will then be stored in a temporary archive 
and sent to the quick-look processing pipeline for system and data quality validation. 
\item After data quality validation, data will be stored in the on-site Observatory Database 
ready for post-processing and the delayed delivery to the off-site 
ASTRI Data Archive in Rome. The user will receive notice of the availability of 
the data and will be able to retrieve these data through a Web-based service and 
will be able to analyze the data using the ASTRI Science Analysis tools. 
\item A central role in the design of the MASS is assigned to the Observation Execution (MASS-OE) 
subsystem, which is responsible for the actual observations requested by the user 
and approved by the ASTRI TAC. This subsystem controls and monitors all ASTRI hardware 
present at the site: telescopes, cameras, auxiliary devices, safety devices, etc. 
It can work automatically but an operator will normally supervise it. 
\end{itemize}

\subsection{The MASS Observation Execution Subsystem}

\begin{figure*}[t]
  \centering
  \includegraphics[width=0.7\textwidth]{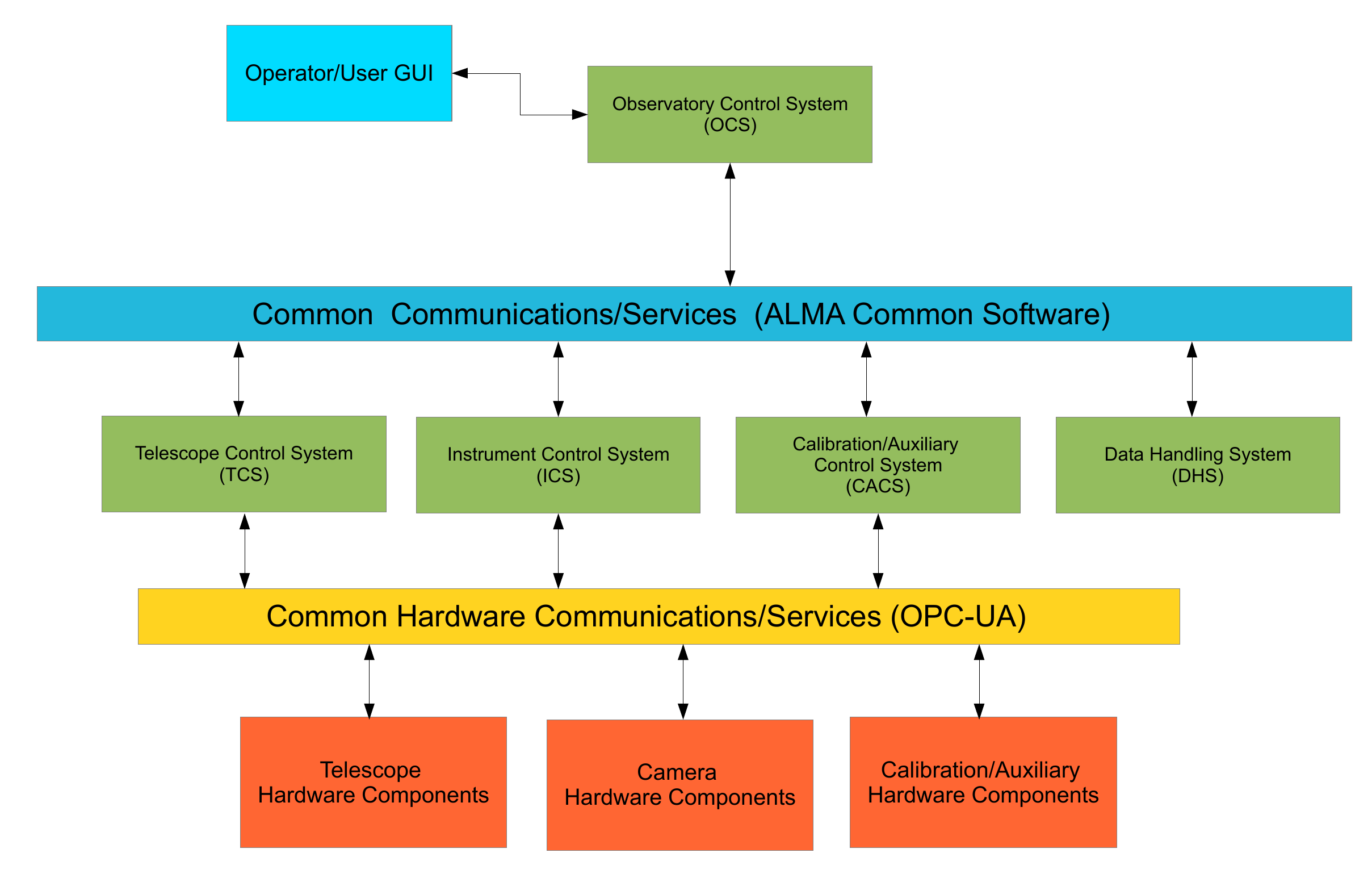}
  \caption{Main Components of the MASS-OE subsystem}\label{fig:fig2}
 \end{figure*}

MASS-OE is a distributed system that, as in many of the most recent telescope control 
system designs, is composed of the following main components (see Figure~\ref{fig:fig2}): 
\begin{itemize}
\item The observatory control system (OCS). 
\item The telescope control system (TCS).
\item The instrument control system (ICS).
\item The data handling system (DHS).
\item The calibration/auxiliary control system (CACS)
\end{itemize}
\subsubsection{The observatory control system}
The OCS is responsible for high-level ASTRI operations; it provides user interfaces, 
scheduling and execution of observations, system monitoring, and coordination of 
the TCS and ICS actions to acquire and monitor bulk data, which then will be stored 
and displayed by the DHS. The OCS also provides high-level graphical user interfaces (GUIs) for system operators 
and observers, together with system and environment status monitors. 

\subsubsection{The telescope control system}

The scope of the TCS is to provide a high quality stable image of a specified point 
on the sky to the ASTRI Camera at the focal plane. The main tasks of the TCS  are coordinating, monitoring  and controlling the activities of its subsystems under instruction from the OCS. The TCS does not include direct 
control of any ASTRI hardware, this is the responsibility of its subsystems, \emph{e.g.},
the Mount Control System and the Active Mirror Control System which will communicate 
with the dedicated OPC (Open Productivity \& Connectivity) -UA (Unified Architecture)\footnote{http://www.opcfoundation.org/} servers installed on the Telescope Control Unit (TCU) and Active Mirror Control Unit (AMCU), the two rack-mounted industrial computers on board the telescope.  The software installed on these two computers is dedicated to the following tasks:
\begin{itemize}
\item command telescope slewing and tracking. 
\item monitor and control the thermal loads on the telescope; 
\item monitor and control the active optics systems; 
\item monitor and control the pointing monitor camera; 
\item monitor and control other simple on/off devices
\end{itemize}
The TCS is not responsible for time-critical operations. All real-time functions are performed at the level of the TCU and the AMCU; this includes all the astronomical coordinate transformations and pointing model corrections needed to produce the commanded position of the telescope at a given time in mount coordinates (\emph{i.e.}, encoder coordinates). In our approach the telescope is a stand-alone system able to manage all the primary functions needed to acquire a celestial source starting from a minimal set of input parameters (\emph{e.g.}, coordinates of the target). In this way high level software should be concerned only with the business logic and not with the details of the hardware operation.

\subsubsection{The instrument control system}

The ICS provides all software components used to configure, control and acquire 
monitoring data from the ASTRI Camera~\cite{camera}. The ICS provides the capabilities for the monitoring and control of the ASTRI Camera
and will allow direct access to all devices controlled by the Detector Control system (DCS) and the Camera Auxiliary 
Devices Control (CADC).

\subsubsection{The data handling system}

The DHS system is responsible for the on-site storing and processing of all the data produced 
by the system (ASTRI Camera bulk data, calibrations, engineering data, housekeeping, etc.). Also, the DHS is responsible for the delivery of the relevant data to the ASTRI off-site archive. A more detailed description of the ASTRI archive design is given here~\cite{data}\\
The ASTRI Camera bulk data are acquired and monitored by the ASTRI Data Acquisition 
system (DAQ). According to the CTA design principle, the data acquisition of the ASTRI Camera is assigned to one 
dedicated computer, the ASTRI Camera Server, in which the Detector DAQ Software (DDS) is deployed and runs. The DAQ is
in charge of collecting the data from the Camera Back End Electronics. A prototype of the DDS software is under development.\\
The data acquired by the DAQ will be processed by the on-site Analysis pipeline which will run the ASTRI Science Analysis Tools developed
taking into account all the peculiarities of the analysis of Cherenkov data. Indeed, 
the Cherenkov data  are more complicated to analyze than data produced by instruments that take 
direct images of the sky. An extensive processing chain is needed to derive the
scientific characteristics of the primary gamma ray from the Cherenkov images of the 
air showers detected by the cameras. The data analysis chain will also  
depend critically on a wide range of parameters related to both the instrument configuration 
and the observing conditions. In preparation for this analysis, a complex set of simulations must also 
be generated to calibrate data in order to estimate both the ``hadroness'' of the incoming 
primary particle and its energy. The Monte Carlo simulation activities ongoing in ASTRI are reported here~\cite{montecarlo}~\cite{montecarlo1}.

\subsubsection{The calibration/auxiliary control system}

The CACS includes several components dedicated to monitor the site condition and provide essential services. Most of these devices will be installed and tested at the Serra La Nave site (see~\cite{sito}), where the ASTRI SST-2M prototype telescope will be installed, 
the following auxiliary instrumentation is part of the CACS and is described in~\cite{sito} :
\begin{itemize}
\item The weather station and the rain sensor
\item The Time Synchronization and distribution system (GPS, etc)
\item The All-Sky Fish-Eye camera
\item The Sky Quality Monitor (SQM)
\item The UVscope-UVSiPM complex
\end{itemize}

\section{The MASS Interprocess Communication frameworks}
From a communications and integration perspective, the MASS-OE is a distributed 
system of semi-autonomous objects interacting with each other through a software communications 
(middleware) backbone. 
 \subsection{The ALMA Common Software framework}
The MASS software will use the ALMA Common Software (ACS) framework{\footnote{http://www.eso.org/\~almamgr/AlmaAcs/index.html}}, developed 
as the foundation for the software system for the Atacama Large Millimeter/submillimeter 
Array (ALMA, see e.g. ~\cite{alma}).  Furthermore, several aspects of CTA (and Mini-Array) system operations 
are nearly the same as those of ALMA. For instance:
\begin{itemize}
\item  the control and synchronization of arrays of telescopes/antennas;
\item the dynamic scheduling of observations; 
\item the acquisition of high-rate data and its storage in petabyte-scale archives;
\item the monitoring, storage and analysis of data from hundreds to thousands of sensors mounted on telescopes  and instruments;
\item the use of scripts to implement diverse observing modes. 
\end{itemize}

ACS, developed by the European Southern Observatory (ESO) for the ALMA project, provides 
the communication and coordination facilities for the main software components. Also it provides
mechanisms for monitoring, logging and alarm management. 
It facilitates distributed development as well as distributed operations through 
a component-container paradigm. Individual developers concentrate on the creation 
and testing of components that, following a standard interface definition and protocol, 
are straightforward to integrate and test with other components in the system. 

Applications are built with ACS as many independent components that communicate 
with each other using a number of common services. Typical technical aspects of 
system integration, such as distribution, deployment and location of other components, 
are hidden from the developers. All these technical issues are the responsibility 
of containers provided by the framework.  

\subsection{The OPC-UA framework}

With respect to the ALMA control software a simplification introduced in the MASS-OE is represented by the use of the standard OPC-UA to access all telescopes devices. This will allow 
us to decouple high-level control software from the specific hardware device used 
and from proprietary communication protocols. 

OPC-UA is designed to be independent of the operating system. It can be deployed 
on Linux, Windows XP Embedded, VxWorks, Macintosh OS X, Windows 7, and classical 
Windows platforms. This also enables OPC to communicate over networks through Web 
Services. Client software is already available that enables access to OPC-UA servers 
by ACS components.

We have performed some tests using the Java OPC-UA console server and client from 
\textit{ProSys} and a Java OPC-UA graphical client from \textit{UaExpert}.  Also we have tested
the integration between ACS and OPC-UA. The 
results of these tests were encouraging and OPC-UA will be used as the standard 
communication protocol to access all ASTRI SST-2M and Mini-Array devices.

\section{Conclusions}
Here we presented the MASS software that will make possible to operate the ASTRI mini-array to perform  the first  scientific observations of a variety of sources (see ~\cite{science}) from the southern CTA site.  The MASS  will be tested on the ASTRI SST-2M prototype that will be installed  at the Serra La Nave site in 2014. 

\vspace*{0.3cm}
\noindent
\footnotesize{{\bf Acknowledgment: }{This work was partially supported by the ASTRI Flagship Project financed
by the Italian Ministry of Education, University, and Research (MIUR) and led by the Italian National Institute of Astrophysics (INAF).
We also acknowledge partial support by the MIUR Bando PRIN 2009.}}

\end{document}